# XPM-forced frequency-oscillating soliton in mode-locked fiber laser

Yudong Cui[1,4], Yusheng Zhang[1], Yunyun Xiang[2], Youjian Song[3] and Xueming Liu[1,*]

*1 State Key Laboratory of Modern Optical Instrumentation, College of Optical Science and Engineering, Zhejiang University, Hangzhou 310027, China*

*2 Second Institute of Oceanography, Ministry of Natural Resources, Hangzhou 310012, China.*

*3Ultrafast Laser Laboratory, College of Precision Instruments and Opto-electronics Engineering, Tianjin University, 300072 Tianjin, China*

*4 cuiyd@zju.edu.cn*

*These authors contributed equally: Yudong Cui, Yusheng Zhang*

*\*corresponding to: liuxm@zju.edu.cn*

**Abstract:** Cross phase modulation (XPM) could induce soliton trapping in nonlinear medium, which has been employed to achieve vector soliton, optical switching and optical analog of gravity-like potentials. These results are generally within the definition in Hamilton system. Here, we report on the observation of a XPM-forced frequency-oscillating soliton (XFOS) whose wavelength exhibits redshift and blueshift periodically like dancing in a mode-locked fiber laser under moderate birefringence. XFOS consists of two orthogonally polarized components exhibiting simultaneous frequency oscillation driven by XPM and gain effect, which allows withstanding higher pulse energy. The pulse trapping is maintained by differentiating the frequency-shift rate. Numerical simulations agree very well with experimental results, revealing an idiosyncratic evolution dynamic for asymmetry pulses in nonlinear dissipative system and envisaging a technique to control pulse feature with preset pulse chirp. XFOS may exist generally in polarization-independent ultrafast lasers, which enriches soliton family and brings useful insights into nonlinear science and applications.

Soliton is described as a wave packet localized in time and/or space that results from a balance between dispersion and nonlinear effects [1][2]. It combined the wave and particle-like behaviors as solitons can cross or collide each other without breaking the integrity [2][3]. The interaction of solitons brings several exciting investigations, such as the formation of vector soliton and the optical analog of event horizon and Hawking radiation, which are dominated by soliton trapping dynamics that is generally caused via cross phase modulation (XPM) mediated by Kerr nonlinearity [4]-[6]. The optical event horizon exists in the formation process of supercontinuum under near-zero dispersion condition and has been used to explain the generation of blue and violet light from infrared femtosecond pump pulses, since the wavelengths of two trapped soliton pulses would be shifted toward shorter and longer wavelength sides, respectively [1][7]. The XPM-induced intensity-dependent nonlinear phase shift would form a gravity-like potential [5][7]. Soliton trapping can maintain the stability by changing the wavelength to accelerate or decelerate pulses, i.e., the relative position mismatch is compensated by dispersion [8]. This effect is employed to control the center wavelength, energy, and duration of signal via a weak control pulse, and also holds the application prospect for optical switching and pulse compression [1][9][10].

Under birefringent conditions, the orthogonally polarized components propagate with slightly different propagation constants that exactly satisfies the condition of soliton trapping [7][11]. Therefore, two polarized solitons can move with a common group velocity, which is called vector soliton known as the typical solution of XPM-modified coupled nonlinear Schrodinger equation (NLSE) [12][13]. Vector soliton directly embodies the soliton trapping effect that two orthogonally polarized solitons exhibit different wavelengths to achieve the synchronization locking [11][14]. All the above ideas are defined in the conservation system where solitons would reach a steady state after a transient change to regain their shape, energy, and velocity [1][11]. Although the similar results can be also reproduced under dissipative conditions, more complicated dynamics is expected in the present of gain and loss.

Solitary wave governed by the Ginzburg-Landau equation (GLE) in far-from-equilibrium nonlinear systems is known as dissipative soliton [15]-[18]. It is helpful to understand the complicated evolution dynamics in fluids, Bose-Einstein condensates, optical systems, and complex networks [18][19]-[21]. Mode-locked fiber laser naturally can provide the appropriate condition for the generation of dissipative soliton [15][22]. Several distinctive dissipative solitons have been investigated in fiber lasers, for instance, dissipative soliton resonance, pulsating soliton, soliton explosion, rogue wave, and some other chaotic behaviors [23]-[26]. By considering the influence of weak birefringence, vector soliton can be formed in mode-locked fiber laser that is insensitive to polarization [27]-[29]. The phase and intensity of two orthogonal polarization components depend on the net cavity birefringence and coupling strength, which can be classified as polarization locked vector solitons (PLVS), polarization rotation vector solitons (PRVS), group-velocity locked vector solitons (GLVS) and so on [28]-[32]. Recently, the interplay and trapping of polarized solitons in micro-resonator have also been concerned [33][34].

The development of ultrafast diagnostic technologies has inspired the latest advances on dissipative soliton [35]-[37]. Sergeyev et al. investigated the polarization dynamics of vector solitons in mode-locked fiber laser and found a new class of slowly evolving vector solitons characterized by a double-scroll chaotic polarization attractor [38]. With dispersive Fourier transformation (DFT) technique, the complex spectrum evolution in terms of the build-up process of soliton [39][40], the dynamics of soliton explosions [41], and the soliton chaotic feature have been resolved experimentally [17][20][42]. Triple-state dissipative soliton that periodically switches over consecutive round trips with quite different spectra was reported by using DFT technique [43]. The single-shot spectral dynamics of vector soliton has also been recorded to show the trapping characteristics [30]. However, the present performance of soliton trapping is not beyond the definition in conservation system, and the birefringence is treated as the perturbation, rather than a dominant factor. In this work, we reported the first observation on a novel soliton operation that pulses spectra exhibit redshift and blueshift alternately in a carbon-nanotube mode-locked fiber laser by means of DFT technique. As it seems to be a typical stable soliton operation with the averaged spectrum, pulse train and radio frequency spectrum, this phenomenon cannot be distinguished without DFT technique. The numerical simulation based on the coupled GLE equation reproduces the experimental results. XPM effect with higher pulse intensity reconstructs the pulse shaping dynamics, which can be called the dissipative vector soliton. However, it is distinct from the previous soliton and vector soliton operations [28]-[32]. These results will bring new understanding about soliton and give vitally new insight into the stable mode-locking operation that will aid in improving laser performance.

**Results**

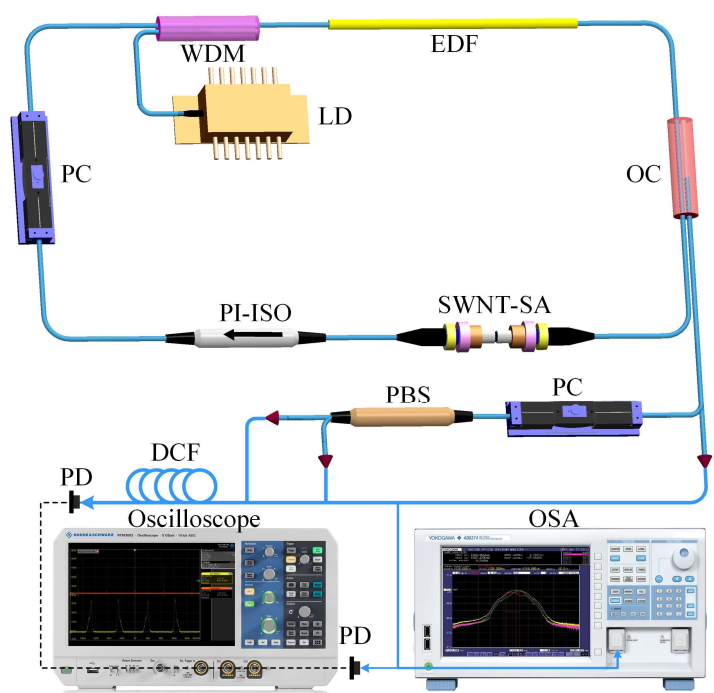

Fig. 1. Schematic diagram of the experimental setup of the mode-locked fiber laser and its ultrafast characterization. EDF: erbium-doped fiber, LD: laser diode, WDM: wavelength division multiplexer, PI-ISO: polarization-independent isolator, OC: optical coupler; SWNT-SA: single-walled carbon nanotube saturable absorber, PC: polarization controller, SMF: single-mode fiber, PD: photodetector, DCF: dispersion-compensating fiber, OSA: optical spectrum analyzer, PBS: polarization beam splitter.

**Experiments.** Figure 1 illustrates the experimental setup and measurement system. The detailed description is shown in Methods section. Mode locking initiates at the pump power of ~10 mW, and could maintain the single-pulse operation until ~14 mW. The soliton operation can be observed with the appropriate state of polarization controller (PC). The exemplary results are measured at ~13 mW and demonstrated in Fig. 2. Figure 2(a) shows the soliton spectral evolution dynamics with respect to the roundtrip. Obviously, the spectra at the odd roundtrip (ORT) and even roundtrip (ERT) display different peak wavelengths and intensity profiles, which can be also seen in Figs. 2(c) and 2(d). The single shot spectra at RT-39 (orange line) and RT-40 (magenta line) in Fig. 2(d) display the asymmetric shape, but their superposition exhibits nearly symmetric Gaussian-shaped profile similar to the OSA-recorded spectrum. With respect to the center wavelength (1560nm) of the averaged spectrum, the spectra for ERT and ORT exhibit redshift and blueshift, respectively. The zigzag line in Fig. 2(a) traces the swing of peak wavelength ($\lambda_p$) whose difference is ~4.3 nm. Moreover, although the optical spectrum changes apparently, the fluctuation of the pulse train in Fig. 2(b) could be almost negligible, which implies the uniformity of pulse energy. The radio frequency spectra of the laser as shown in Supplementary Fig. 1 further verify the stability of pulse train [44]. It seems like the same to that of traditional stable mode-locking operation [22][28][45]. However, the spectral dynamics has not been reported ever before, and is completely distinct from the dynamics of dissipative soliton and the derivative states [1][4][15]-[20][22]-[32].

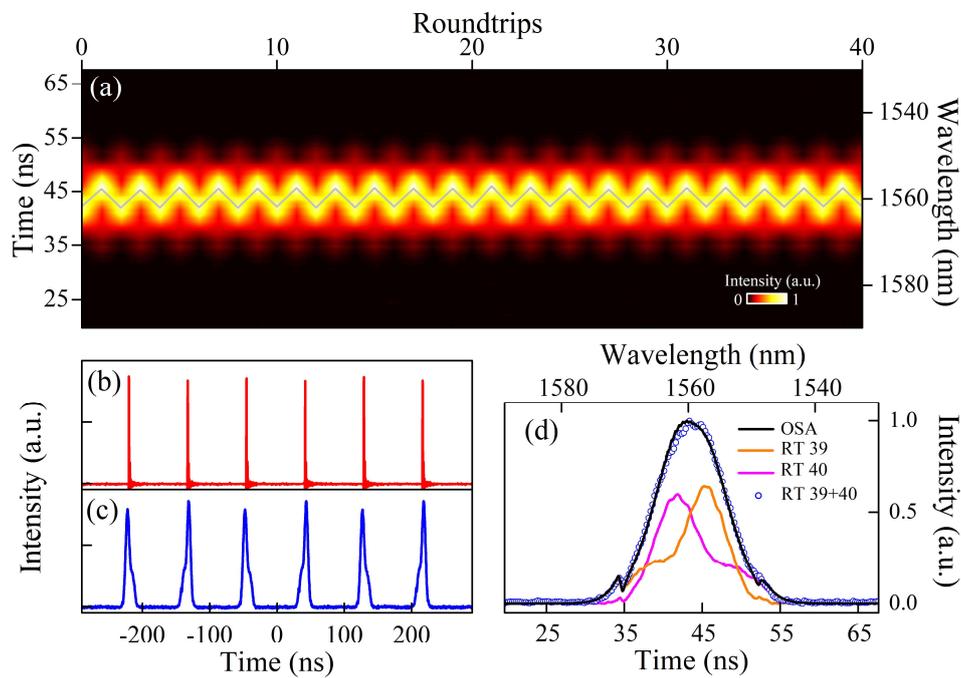

Fig. 2. Experimental results in mode-locked fiber laser. (a) Real-time observation of the evolution dynamics. The recorded time series is segmented with respect to the RT, and the intensity profile evolves along with time (vertical axis) and roundtrips (horizontal axis). Experimental measured pulse trains recorded by oscilloscope without (b) and with (c) DFT technique. (d) Optical spectra measured by OSA and DFT process. The single-shot spectra correspond to the last two frames in (a).

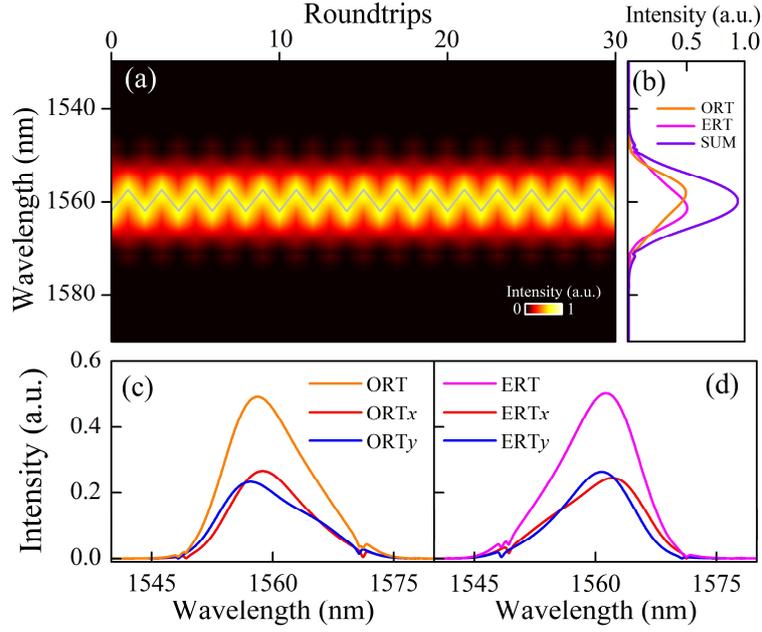

Fig. 3. Numerical simulations based on the cavity round-trip model. (a) Evolution dynamics of soliton within 30 roundtrips. The zigzag line traces $\lambda_p$. (b) Spectral profiles for the last two roundtrips. *x*-axis and *y*-axis components in the simulation for (c) odd roundtrip (ORT) and (d) even roundtrip (ERT).

**Numerical simulations.** To confirm the experimental observations, the laser cavity is described with the cavity round-trip model to consider every action of the cavity components [32][45]. The coupled Ginzburg-Landau equations (CGLEs) are employed to calculate the pulse propagation in the weakly birefringent fibers [1][29][32]. In our model, the terms of group-velocity dispersion (GVD), self-phase modulation (SPM), and XPM are included. However, an asymmetric solution is obtained, although no high-order term is considered. The simulation model is shown in Methods section in detail. Different from the previous results that the pulse evolution is self-consistent in one round [32], pulses in our simulation evolve with the period of two roundtrips. So the spectra of output pulses over two consecutive rounds display the distinct features as shown in Fig. 3(a), which is almost exactly the same as the experimental results in Fig. 2(a). The two-dimensional contour plot is characterized by a zigzag contrail. Fig. 3(b) shows the spectral profiles of two consecutive rounds which shift toward shorter and longer wavelength respectively and their superposition is symmetrical. The qualitative features of the simulation dynamics are roughly in accordance with the experimentally observed spectral shift in Fig. 2. However, the spectral profiles, peak positions and intensities have obvious differences by comparison with the experimental results in Fig. 2(d), which may be caused by the higher-order dispersion and higher-order nonlinear effects. The simulation results would be a steady solution, because it still follows the regular evolution when the simulation is operated for tens of thousands rounds. As the simulation based on the CGLE can nearly reproduce the experimental results, it indicates that the phenomenon originates from the birefringence in the fiber laser leading to enhanced XPM. As a result, the novel phenomenon is named as XPM-forced frequency-oscillating soliton (XFOS).

In fact, XFOS for each round consists of two orthogonal polarization components (along $x$ and $y$ axes in the simulation) which is displayed in Figs. 3(c) and 3(d). Two polarization components have different wavelengths and intensities, but both show the similar evolution dynamics with that in Figs. 2 and 3(a) (also see Supplementary Figs. 2(a) and 2(b)). This means that $x$-axis and $y$-axis components red-shift or blue-shift simultaneously, which is hardly explained by the traditional mechanism of soliton trapping in the birefringent fiber [2][30][31][34]. With the aid of PBS and PC, pulses are split into two orthogonal polarization components in the experiment, and the similar results could be obtained (see Supplementary Figs. 2(c) and 2(d)). These results further verify the numerical simulation. Nevertheless, the dynamics mechanism of XFOS cannot be understood with the output pulse train in Figs. 2 and 3, and it is also not possible to obtain intracavity characteristics experimentally by prior art. The intracavity evolution can only be speculated with numerical simulations, which should be credible as the simulation results are highly consistent with the experimental results.

The intracavity spectral evolutions of XFOS along with $x$-axis and $y$-axis at two consecutive rounds are exhibited in Figs. 4(a) and 4(b). The two rounds are denoted as ORT and ERT. Evolution of the corresponding $\lambda_p$ and bandwidth ($\Delta\lambda$) are drawn in Figs. 4(c) and 4(d), respectively. The pulse width, energy and the chirp evolution of two orthogonal components are shown in Supplementary Fig. 3. In ORT, pulses are firstly amplified in EDF and the spectrum firstly becomes narrow and then broadens, while the pulse is always compressed reaching the minimum after SA. The process is distinct from the general pulse evolution under the dispersion-managed condition [22]. As shown in Fig. 4(e), the time-bandwidth product (TBP) at the position of around 5 m is the smallest in EDF, implying that the negative linear chirp ($C_L$) has been compensated and pulses begin to be endowed positive $C_L$. But pulses can continue to be compressed owing to the remained nonlinear chirp ($C_{NL}$), and TBP increases actually with the broadening spectrum. Meanwhile, obvious blue shift and energy exchange can be observed for two orthogonal components. After the loss of SA, there is still a few energy exchange and the spectra begin red-shifting. Pulses are stretched by the dispersion of SMF, then entering the next roundtrip. In ERT, the variation of temporal width, energy and spectral width is similar to these in ORT, while it is reversed for $\lambda_p$ exhibiting redshift before SA and blueshift after SA. The temporal and spectral profiles of XFOS swing like dancing during two roundtrips. The soliton dance is vividly illustrated in the animation of Supplementary Movies 1 and 2. The relative peak position ($t_p$) in Fig. 4(f) is to illustrate the dynamical pulse trapping. Noting that the temporal separation of two orthogonal polarized pulses is no more than 100 fs in the whole process and is much shorter than the pulse width, indicating that the two pulses can trap each other in the cavity, although the separation changes slightly owing to the influence of dispersion and birefringence.

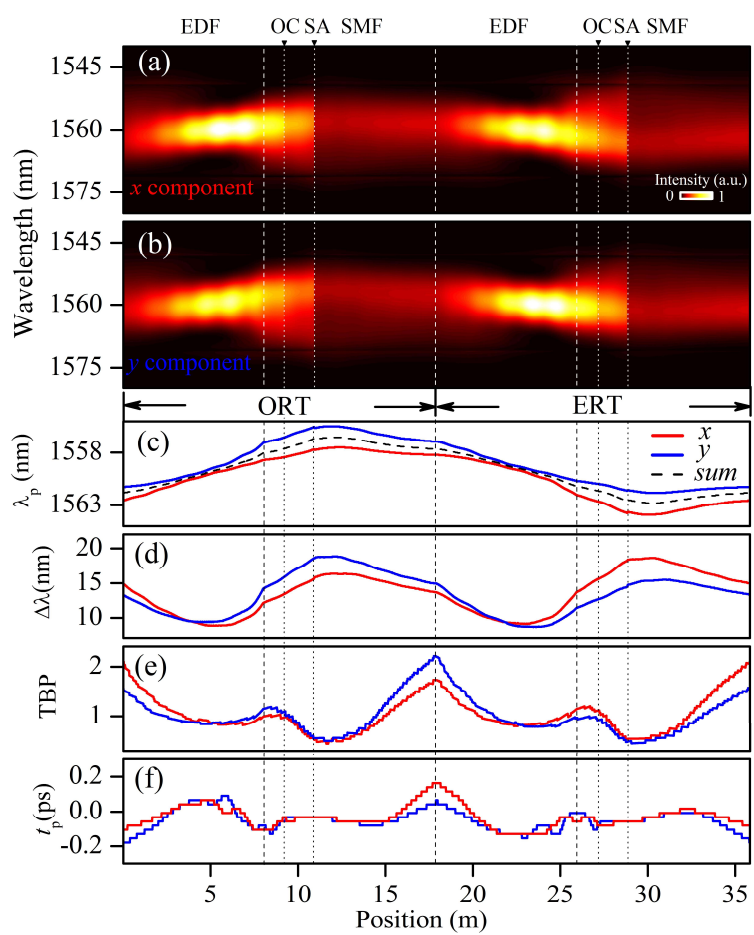

Fig. 4. Intracavity spectral evolution dynamics of soliton over two consecutive roundtrips for (a) $x$ component and (b) $y$ component. Here the two consecutive roundtrips are denoted as ORT and ERT. Evolution of (c) spectral peak ($\lambda_p$) and (d) bandwith ($\Delta\lambda$) for the two orthogonally polarized

components and their superposition. (e) Evolution of time-bandwidth product (TBP) for two orthogonal components. (f) Evolution of pulse peak positions ($t_p$) for the two orthogonal polarization components. The jagged line results from the time resolution in the simulation.

**Discussion**

The mechanism of traditional vector soliton is discussed firstly. For GLVS which exists under suitable birefringence, two orthogonally polarized components could trap each other against the birefringence by shifting their spectra to shorter and longer wavelengths, while the whole spectrum (superposition of two polarization) normally maintains the symmetrical shape [1][27][32]. GLVS can also be obtained with the theoretical model of this work, when the gain saturation energy is set as $E_s$=25 pJ (see Supplementary Fig. 4). By comparison with XFOS, the evolution dynamics is obviously distinct that the frequency shift of two orthogonally polarized pulses is in the opposite rather than the same direction. It is also different from the conventional vector soliton that is reported in Ref. [1], which should be attributed to the periodical stretch and compression of asymmetric spectra and pulses, the typical characters under dispersion-managed condition [22][46]. The spectral variation depends on the fact that a negatively chirped pulse undergoes spectral narrowing through SPM (also known as inverse four-wave mixing [47]) and spectra exhibits broadening with the initial positive frequency chirp [1]. However, it is still a conservation process under the periodically changed dispersion.

The evolution dynamics of XFOS also follows the similar rules, but is with a new balance of GVD, SPM, XPM, gain and loss, which can be explained as follows with the schematic diagram in Fig. 5(a) and the typical results of chirp in Supplementary Fig. 5. SPM induces the negative and positive $C_{NL}$ along the pulse leading and trailing edge resulting in the down-frequency and up-frequency shift, respectively. The new-generated wavelength depends on the pulse chirp, i.e., frequency distribution in the temporal domain. Hence, the spectra are narrowed with negative $C_L$ (at 0 m~5.1 m and 29.8 m~35.8 m, 12.2 m~22.2 m in Fig. 4(d), corresponding to (i) and (v) in Fig. 5(a)) and broadened with positive $C_L$ (at 5.1 m~12.2 m, 22.2~29.8 m in Fig. 4(d), corresponding to (iii) and (vii) in Fig. 5(a)). During the processes, the off-center peak wavelengths would be shifted because of the asymmetric spectral profile originating from XPM. For pulses in (i) or (iii), the spectral peak possesses longer/shorter wavelength that is located at the trailing edge so that it would show blueshift. For pulses in (v) or (vii), the spectral peak possesses shorter/longer wavelength that is located at the leading edge so that it would show redshift. The nonlinear phase shift for two linear polarization components ($\varphi_x$ and $\varphi_y$) can be expresses as $\varphi_x=\gamma(P_x+2/3P_y)L$ and $\varphi_y=\gamma(P_y+2/3P_x)L$, respectively. $P_x$ and $P_y$ are the peak power of the two components, respectively, and $L$ is the cavity length. $C_{NL}$ is $\delta\omega_x(T)=\partial\varphi_x/\partial T=\gamma L(\partial P_x/\partial T+2/3\partial P_y/\partial T)$ and $\delta\omega_x(T)=\partial\varphi_y/\partial T=\gamma L(\partial P_y/\partial T+2/3\partial P_x/\partial T)$. Consequently, XPM can introduce the asymmetric nonlinear phase when two polarized pulses overlap incompletely, which leads to the formation of the asymmetric pulse and spectrum. During the evolution of XFOS, the frequency-shift rate will be modified by the XPM to accelerate the back pulse and decelerate the front pulse. On the other hand, XPM can also induce the nonlinear phase with the same sign as SPM to enhance SPM effect, which can be seen from Fig. 4 and also shown in Supplementary Fig. 6. During the spectral narrowing process, the asymmetrical spectra gradually become more symmetrical as $\lambda_p$ approaches 1560 nm. And it can also be believed that the XPM-induced asymmetry can be neutralized by SPM under negative $C_L$. During the spectral broadening, the asymmetry is boosted by SPM under positive $C_L$, as $\lambda_p$ shifts away from center wavelength.

When $C_L$ is near zero, the wavelength distribution along temporal domain is dominated by $C_{NL}$, and the evolution dynamics is more complicated. SPM-generated frequency components will broaden the spectrum for unchirped pulses [1]. The frequency shift can be compensated when pulses possess a slight negative $C_L$. Owing to the combined action, the spectral bandwidth changes slightly during the transition from broadening to compression. However, the blueshift/redshift of pulses will be continued for the cases of (ii)/(vi), because the long/short wavelength components within the pulse trailing/leading edges have higher energy. The actually spectral variation is related to the shape of chirp which is demonstrated in Fig. 5(a). For instance, around 5 m in Fig. 4 corresponding to (ii), the spectra continue to be blueshift without obvious variation of bandwidth. After the point with zero $C_L$, the spectra can still exhibit blueshift, which is completely different from that of GLVS. It is found that the process is dependent on the pulse intensity. The propagation of pulses with different intensities is modelled by NLSE without gain and XPM effect (see Methods for details). As shown in Fig. 5(b), pulses with low intensity (from $0.6I_0$ to $1.0I_0$) exhibit blueshift firstly and then turn into redshift, while $\lambda_p$ of pulses with high intensity ($2I_0$) always shifts to the shorter wavelength. High intensity can force $\lambda_p$ to be shorter than 1560 nm–before spectra begin to broaden. Under the circumstances, the spectral tilt direction indicating the spectral energy distribution is changed, and then $\lambda_p$ exhibits blueshift with spectral broadening. Pulses with low intensity would not change the spectral vergence so that $\lambda_p$ that is longer than 1560 nm shows redshift since more energy is located at trailing edge, which results in GLVS as shown in Supplementary Fig. 4. The corresponding spectral evolutions are shown in the Supplementary Fig. 7. For the cases from $1.2I_0$ to $1.7I_0$, they are the intermediate states. Although $\lambda_p$ is shifted shorter than 1560 nm, it could change to be redshifted finally, because the spectral vergence is not changed and the energy is partial to longer wavelength. However, the spectral profiles have been deviated from the general asymmetric pulse shape. It should be noted that the underlying dynamics is similar for (vi). XPM and gain both can enhance the nonlinear effect inducing the change of spectral tilt direction, as shown in Fig. 5(c) and Supplementary Fig. 6(b). Actually, to maintain the self-consistent of the evolution, $\lambda_p$ should be just around 1560 nm with a nearly symmetrical spectrum when the spectrum is the narrowest. Higher nonlinear effect or pulse intensity will break the operation stability. The shifting rate of two polarized pulses can also be changed by XPM. $\lambda_p$ of two orthogonal polarized pulses would firstly approach each other, and then move away from each other. For the cases of (iv) and (viii), the spectral tilt direction would not be changed and the shift direction of $\lambda_p$ is revered accompanying with the slightly changed bandwidth.

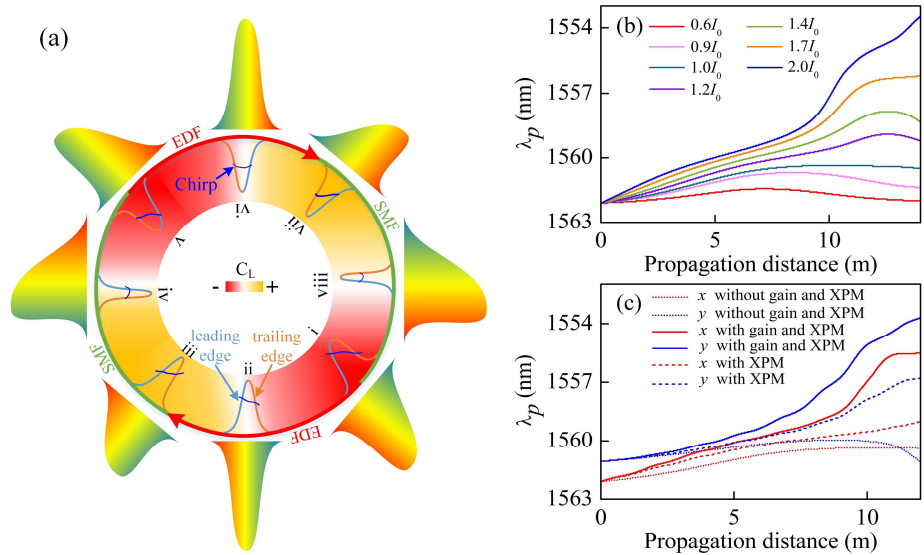

Fig. 5. (a) Illustration of the evolution dynamics of XFOS in two consecutive roundtrips. The middle line denotes periodically changed EDF and SMF; the outer and inner layer shows the spectral and temporal evolution, respectively; the background color shows the change of $C_L$; the blue lines denote the chirp; the different color of pulses is on behalf of leading and trailing edges. The process is divided into eight typical parts from (i) to (viii). (i) At the beginning of EDF in ORT, $C_L$ is negative. (ii) Near the point of zero $C_L$ in EDF. (iii) After the zero point, $C_L$ becomes positive. (iv) Near the point of zero $C_L$ in SMF of ORT (v) Then $C_L$ becomes negative. (vi) Near the point of zero $C_L$ in EDF of ERT. (vii) After the zero point, $C_L$ becomes positive. (viii) Near the point of zero $C_L$ in SMF of ERT. (b) Evolutions of $\lambda_p$ with different pulse intensity. (c) Evolutions of $\lambda_p$ for three cases: including XPM, both including XPM and gain, and neither of them is included.

The difference of the formation condition for XFOS and GLVS is the pulse energy depending on the intracavity gain and loss. The larger pulse energy brings the larger nonlinear phase shift driving the wavelength to oscillate in a larger range. It should be noted that the frequency shift can reduce the nonlinear phase so that pulses can bear higher pulse energy. However, the wavelength difference of two polarized pulses is related to the linear birefringence, which can be more strictly restricted to the specific value by the larger pulse energy, i.e., the wavelength difference varies in a smaller range. To understand the formation of XFOS, it can be analogous to the forced oscillation of two vibrators connected with a spring. When the driving forces on two vibrators are nearly the same and within the magnitude and range of the elastic force of spring (weak coupling between two polarized solitons), two vibrators can oscillate simultaneously via moving toward the opposite directions, and keep the center of the spring still, like the process of GLVS in Supplementary Fig. 4. When the elastic force of spring is much larger (strong coupling) and the range of driving forces also extend, the driving force hardly resonates with the spring. Slight difference of two driving forces would force two vibrators to oscillate as a whole in a larger range. Meanwhile, the distance of two vibrators will also change periodically. This process is similar to the coupled dissipative parametric resonator that can be used to elucidate the many-body time crystal [48]. Here, the results of two coupled solitons also show the period-doubling bifurcation revealing the possibility for the realization of time crystals in optics.

Seemingly, the results that the frequencies of two interacted soliton shift to the same direction are in conflict with the rule of energy conservation. The driving forcing of frequency oscillation originates from SPM and XPM which is a degenerate Four-wave mixing process [1], accompanied with the variation of spectral profile and intensity. In fact, the frequency shift of two polarized solitons is not synchronous, but their separation would change periodically reflecting the soliton interaction. Some other effects also play important roles in the dynamics of XFOS, such as saturable absorption of SWNT-SA, gain competition and gain filtering effect. In fact, it's hard to explain every detail of the complex process. XFOS cannot be formed without dissipative conditions, which is the dynamic equilibrium dominated by XPM in a dissipative system. Noting that the self-consistent evolution depends on the periodical variation of pulse energy, XFOS would deviate from the general evolution path. This indicates the inherent dissipative characters of XFOS. Thus, XFOS can be regarded as a kind of dissipative vector soliton. The swing of $\lambda_p$ in one round maintains the identical roundtrip time for two successive rounds, although the observed adjacent pulses from Fig. 2 have the different peak wavelengths, which challenge the stability of longitudinal mode. However, some symmetry beauty is included in the evolution of asymmetric pulses at two rounds, such as the center wavelength and bandwidth of $x$-axis component at ORT and $y$-axis component at ERT. The spectral superposition of two polarized pulses is nearly asymmetric along 1560 nm, the center wavelength of the average spectrum in Fig. 2. Thus the superposition of two output adjacent pulses is almost symmetric that is similar to the typical soliton operation.

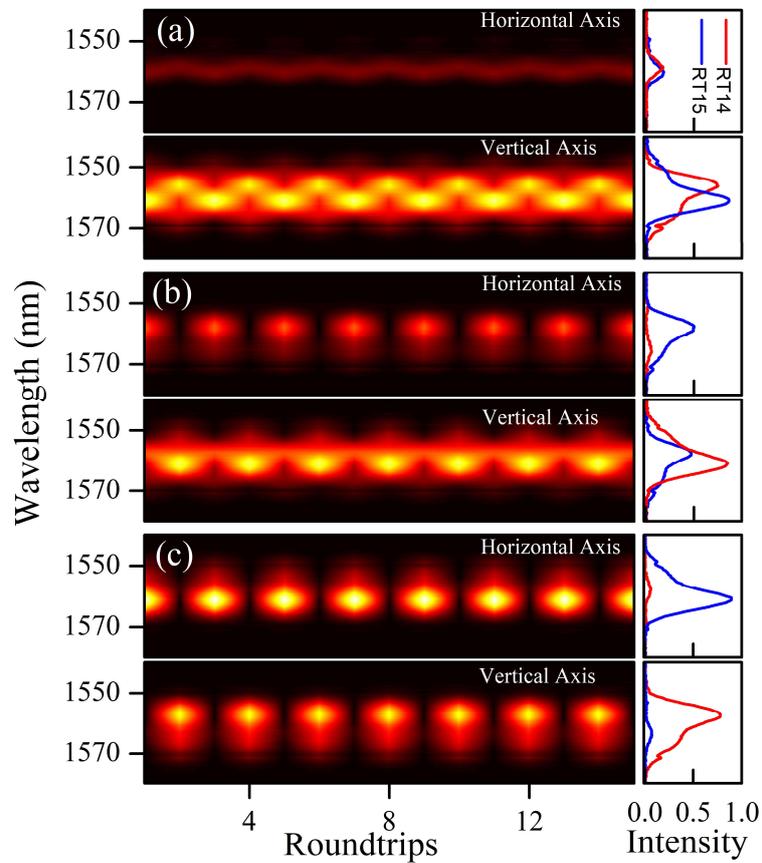

Fig. 6. Polarization evolution dynamics for three different cases. The polarization directions of two

branches of PBS are denoted as horizontal axis and vertical axis, respectively. (a) Pulses both concentrate on vertical axis. (b) Red-shifted pulses concentrate on vertical axis and blue-shifted pulses are divided into two parts. (c) Red-shifted and blue-shifted pulses concentrate on the different axes.

The polarization features of XFOS are examined by using a PBS that can split pulses into two orthogonal polarization components. The polarization angle of pulses would be adjusted by the PC before PBS, inducing the different energy distributions between two branches, but the relative polarization relationship of two adjacent pulses would not be changed. The polarization evolution dynamics can be inferred like that for the studies of PRVS [28][30]. Figure 6 demonstrates the typical results for three different cases with their pulse trains shown in Supplementary Fig. 8. In the case of Fig. 6(a), the majority of energy of all pulses can be concentrated on the vertical axis, which means that the polarization of all pulses is nearly parallel. The pattern in Fig. 6(b) implies that when the red-shifted pulses are concentrated on the vertical axis, the blue-shifted pulses will have to be split into two branches. As a result, the polarization of the adjacent pulses must have an angle. In fact, the splitting proportion can be changed by adjusting the intracavity PC, and Fig. 6(c) is the limiting case where red-shifted and blue-shifted pulses concentrate on the different branches. In this case, the polarization of adjacent pulses is nearly perpendicular. It seems to contradict the results in Supplementary Fig. 2 which is achieved with the different states of the PC before PBS. The distinct polarization evolutions can be explained by the different phases between the two orthogonal components. The coherent superposition of the two orthogonal components in Supplementary Fig. 2(a) would yield the distinct polarizations. In this way, XFOS is further distinguished from vector soliton.

In conclusion, a XFOS is demonstrated in a dissipative system under weak linear birefringence by using DFT technique. Pulses would exhibit redshift and blueshift for two consecutive rounds, respectively, while their spectral superposition, pulse train and radio frequency have the similar features with that of traditional soliton operation. The relative angle of polarization directions of adjacent pulses can be tuned from orthogonal to parallel by adjusting the intracavity PC. In previous works, XFOS may be regarded as PLVS for Fig. 6(a) and PRVS for Figs. 6(b) and 6(c), which should thank the DFT technique for the capability of resolving the detail of transient periodical signal. With the modelling based on the coupled GLE equation, the experimental results can be reproduced numerically, which verifies the speculation that XPM dominates the pulse-shaping dynamics. The dancing performance is explained with the variation of the chirp, revealing a novel technique to control the wavelength of ultrashort pulse. The results illustrate the new possibility of holding the dynamic equilibrium of XPM, SPM and dispersion in dissipative system, and are helpful to the laser design and the understanding of ultrafast laser.

**Methods**

**Experimental setup.** A section of about 8 m erbium-doped fiber (EDF) is utilized as the gain medium, and a 976 nm laser diodes (LD) provides pump power via a 980/1550 nm wavelength division multiplexer (WDM). A polarization-independent isolator (PI-ISO) ensures the unidirectional operation. A 10% port of a fused 90/10 optical coupler (OC) is used to output pulses. The mode locker is fabricated with single-walled carbon nanotube saturable absorber (SWNT-SA), whose optical response is insensitive to the polarization. A polarization controller (PC) is employed to change the linear birefringence. The other fibers in the cavity are the standard single-mode fiber (SMF). The dispersions of SMF and EDF are -21.7 $ps^2$/km and 20 $ps^2$/km at 1.56 μm, respectively, resulting in a net cavity dispersion of about −0.05 $ps^2$. The total cavity length ∼17.9 m entails a fundamental repetition rate of 11.46 MHz.

The output pulses are monitored with a conventional OSA and a real-time digital oscilloscope together with high-speed photodetector (PD), respectively. The DFT technique is implemented by temporally stretching solitons in a 5-km-long dispersion-compensating fiber (DCF) with the dispersion of about −160 ps/nm/km. A part of the output is firstly measured by OSA, DFT process and PD to obtain the time-averaged spectrum, single shot spectrum and pulse train, respectively. The rest is divided into two orthogonal polarization components by a polarization beam splitter (PBS), and then detected via DFT process.

**Numerical simulations.** To confirm the experimental observations, the laser cavity is described with the cavity round-trip model to consider every action of the cavity components [32][45]. The coupled Ginzburg-Landau equations (CGLEs) are employed to calculate the pulse propagation in the weakly birefringent fibers [1][29][32][49].

$$\frac{\partial u_x}{\partial z} = i\Delta\beta u_x - \delta\frac{\partial u_x}{\partial t} - i\frac{\beta_2}{2}\frac{\partial^2 u_x}{\partial t^2} + i\gamma\left(|u_x|^2 + \frac{2}{3}|u_y|^2\right)u_x + \frac{i\gamma}{3}u_y^2 u_x^* + \frac{g}{2}u_x + \frac{g}{2\Omega_g}\frac{\partial^2 u_x}{\partial t^2}, \quad (1a)$$

$$\frac{\partial u_y}{\partial z} = -i\Delta\beta u_y + \delta\frac{\partial u_y}{\partial t} - i\frac{\beta_2}{2}\frac{\partial^2 u_y}{\partial t^2} + i\gamma\left(|u_y|^2 + \frac{2}{3}|u_x|^2\right)u_y + \frac{i\gamma}{3}u_x^2 u_y^* + \frac{g}{2}u_y + \frac{g}{2\Omega_g}\frac{\partial^2 u_y}{\partial t^2}. \quad (1b)$$

Here, $t$ and $z$ are time and propagation distance, respectively. $u_x$ and $u_y$ donate the slowly varying envelopes of two orthogonal polarized pulses. $2\Delta\beta=2\pi\Delta n/\lambda$, $2\delta=2\Delta\beta\lambda/2\pi c$, and $\Delta n$ are the wave-number difference, inverse group velocity difference, and refractive index difference between two polarization modes, respectively. Here assuming all the fiber has the same birefringence, considering the fact that the birefringence is randomly distributed along the fiber and the polarization results from the accumulation appearance for all the intracavity fiber. $\lambda$ is the operating wavelength, and $c$ is the speed of light in vacuum. $\beta_2$ is the second-order dispersion coefficient, and $\gamma$ is the cubic refractive nonlinearity of the fiber. $\Omega_g$ is the bandwidth of the laser gain. $g$ is the saturable gain of the fiber and is set as zero for SMF. For the EDF, the gain saturation is described as $g=g_0\exp(-E_p/E_s)$, where $g_0$, $E_p$, and $E_s$ are the small-signal gain coefficient, pulse energy, and gain saturation energy, respectively. The saturable absorber is modeled by a transfer function with transmission of $T=1-\alpha_0-\alpha/(1+P/P_{sat})$, where $\alpha_0$ is the unsaturated loss, $\alpha$ is the modulation depth, $P$ is the instantaneous pulse power, and $P_{sat}$ is the saturation power.

The CGLEs are solved with a standard split-step Fourier technique that converges to stable solutions. We use the following parameters in our simulations to possibly match the experimental conditions. $g_0$=3 dB/m, $E_s$=55.5 pJ, $\Omega_g$=40 nm, $\gamma$=4.5 W$^{-1}$km$^{-1}$ and $\beta_2$=20 ps$^2$/km for EDF; $\gamma$=1.3 W$^{-1}$km$^{-1}$ and $\beta_2$=−21.7 ps$^2$/km for SMF. The above parameters are obtained from the data sheets of fiber products. The total length of EDF and SMF are 8 and 9.9 m, respectively. $\lambda$=1560 nm, $c$=3×10$^8$ m/s; $\alpha$=0.1, $\alpha_0$=0.5, $P_{sat}$=50W; The fiber birefringence $\Delta n$ is set as 5×10$^{-7}$ for all fibers. The initial condition used in our numerical simulation is a weak Gaussian pulse.

To compare the evolutions of pulse peak wavelength with different intensity, the conventional nonlinear Schrödinger equation (NLSE) is used to calculate the propagation of pulses. Here the effects of gain and XPM are excluded.

$$\frac{\partial u}{\partial z} = -i\frac{\beta_2}{2}\frac{\partial^2 u}{\partial t^2} + i\gamma|u|^2 u, \quad (2)$$

where $u(z, t)$ is the amplitude envelope of the pulse field, $\beta_2$ is the dispersion of fiber and $\gamma$ is the cubic refractive nonlinearity of the fiber. Here the used parameters are the same with EDF. We assume the *x* component of the pulse at position=0.01 m as the initial pulse (the temporal intensity profile is shown in Supplementary Fig. 6(a)). The peak intensity of the initial pulse here is denoted as $I_0$. The results are shown in Fig. 5(b).

The nonlinear phase shift can be enhanced by gain and XPM. The evolutions of peak wavelength for the cases with XPM, with XPM and gain, without XPM and gain are calculated. Equation (1) is utilized to model the propagation with XPM. When gain is not considered, $g_0$ is set as 0. Equation (2) is utilized to model the propagation without XPM and gain. The x component of the pulse at position=0.01 m is assumed as the initial pulse.

## Acknowledgements

The authors acknowledge support from the National Natural Science Foundation of China under Grant Agreements 61705193, 41506140, 11774310, 61525505, 61975144, 61805212, and the Fundamental Research Fund of Second Institute of Oceanography, Ministry of Natural Resources (JG1719).

## Author contributions

Y.D.C. performed the experiments and the data analysis. Y.D.C. and Y.S.Z. wrote the most of the manuscript and supplemental material, and performed the numerical simulations. Y.D.C., Y.S.Z., Y.Y.X., Y.J.S. and X.M.L. contributed to the fruitful discussions and the write-up. Y.D.C. conceived the project and directed the work.